\documentclass[twocolumn,table,xcdraw]{aastex631}
\usepackage{hyperref}
\usepackage{nameref}
\usepackage{orcidlink}

\usepackage{silence}
\usepackage{url}
\usepackage{booktabs, multirow}
\usepackage{amsmath}
\usepackage{hyperref}

\newcommand{\sgn}{\mathrm{sgn}}
\newcommand{\ee}{\end{equation}}
\newcommand{\be}{\begin{equation}}
\newcommand{\bea}{\begin{eqnarray*}}
\newcommand{\eea}{\end{eqnarray*}}
\newcommand{\ba}{\begin{array}}
\newcommand{\ea}{\end{array}}
\newcommand{\p}{\partial}

\renewcommand{\d}{{\mathrm{d}}}
\newcommand{\e}{{\mathrm{e}}}
\renewcommand{\i}{{\mathrm{i}}}
\newcommand{\bw}{{\mathbf{w}}}
\newcommand{\F}{{\cal F}}
\newcommand{\bJ}{{\mathbf{J}}}
\newcommand{\br}{{\mathbf{r}}}
\newcommand{\bbv}{{\mathbf{v}}}
\newcommand{\cc}{{\textsf{c}}}

\begin{document}

\title{A generalized linear matrix method for normal modes in collisionless stellar disks}

\author[0000-0002-4596-1222]{Evgeny V. Polyachenko}
\affiliation{SnT SEDAN, University of Luxembourg \\
29 boulevard JF Kennedy, L-1855 Luxembourg, Luxembourg}
\affiliation{Institute of Astronomy, Russian Academy of Sciences,\\
48 Pyatnitskaya st, Moscow 119017, Russia}
\email{evgeny.polyachenko@uni.lu, epolyach@inasan.ru}

\author[0000-0002-6060-694X]{Ilia G. Shukhman}
\affiliation{Institute of Solar-Terrestrial Physics, Russian Academy of Sciences, Siberian Branch \\
P.O. Box 291, Irkutsk 664033, Russia}
\email{shukhman@iszf.irk.ru}


\begin{abstract}
   We generalize the linear matrix method for computing normal modes in collisionless stellar disks to distribution functions with sharp edges at zero angular momentum ($L=0$). The generalization adds boundary-integral terms to the matrix equation without increasing its size. We validate the method by computing $m=2$ modes for two Kuzmin--Toomre disk models (Miyamoto $n_{\rm M}=3$ and Kalnajs $m_{\rm K}=6$ families) and comparing the eigenvalues with those obtained from an independent nonlinear matrix method based on logarithmic-spiral expansions. A systematic convergence study over grid resolution and harmonic truncation yields eigenvalues accurate to ${\sim}\,0.003$ in both pattern speed and growth rate. Unlike the nonlinear method, the linear method naturally incorporates gravitational softening, enabling the computation of eigenmodes for softened disk models. The implementation in Julia with GPU acceleration is openly available.
\end{abstract}


\section{Introduction}
\label{sec:introduction}

The Toomre stability criterion $Q > 1$ for flat stellar disks was derived under the assumption that spiral patterns are tightly wound and that stellar orbits are nearly circular \citep{LinShu1964, Toomre1964}. The theory assumed tight winding but allowed perturbations of arbitrary azimuthal number $m$; the criterion itself was obtained in the limiting case $m = 0$, i.e.\ for purely axisymmetric perturbations. Early $N$-body experiments demonstrated, however, that disks satisfying $Q > 1$ remain susceptible to large-scale bar-like instabilities \citep{1970ApJ...161..903M, 1971ApJ...168..343H, 1973ApJ...186..467O}. \citet{1997JETP...85..417P} showed that for realistic disk models complete stabilisation requires values of the Toomre parameter as large as $Q \gtrsim 3$ \citep[see also][]{1997AstL...23..525P}.

The first analytical tool capable of capturing disk instabilities without restricting consideration to nearly circular orbits or tightly wound spirals is the matrix method introduced by \citeauthor{1971ApJ...166..275K} (\citeyear{1971ApJ...166..275K}, \citeyear{Kalnajs77}). The method requires two relations between the perturbed surface density and the perturbed gravitational potential in order to formulate an eigenvalue problem for the complex mode frequency $\omega$ and the corresponding eigenvector describing the shape of the perturbation. The first relation is provided by the linearized collisionless Boltzmann equation; the second is the Poisson equation, which can be replaced by an expansion in a biorthonormal basis set of surface density--potential pairs. In its original formulation the method was cast in Lagrangian form \citep{Kalnajs77} and is sufficiently complex that it has not been widely adopted.

A more formal re-derivation of the Kalnajs matrix method was later carried out by \citet{Hunter92} and \citet{PC97}, who obtained an Eulerian form of the matrix elements. The Lagrangian and Eulerian formulations of the matrix equation are equivalent provided that the distribution function vanishes at the boundary of the accessible region of action space. We refer to such distribution functions as \emph{regular}, since for them the standard (unmodified) matrix method applies without additional boundary terms. When this condition is not satisfied (for example, in unidirectional disks whose distribution functions are discontinuous at zero angular momentum), the Eulerian formulation requires extra boundary-integral terms, the omission of which, as shown by \citet{JH05}, can lead to substantial errors.

The alternative matrix method proposed by \citeauthor{Pol05} \citep{2004MNRAS.348..345P, Pol05} is also based on action–angle variables and, like the Kalnajs approach, does not impose restrictions on orbit shape or spiral winding. Its advantages are that (i) it takes the form of an ordinary linear eigenvalue problem, (ii) does not require choosing a surface density–potential pairs, and (iii) includes an explicit gravitational softening parameter that enables construction of softened models. The last feature is particularly important: since $N$-body simulations of stellar disks typically use gravitational softening, a meaningful comparison of simulated and analytical growth rates requires computing eigenmodes with the same softening. The method’s main drawbacks are the strict constraints on the phase‑space grid parameters and the number of the Fourier harmonics of the perturbed DF and potential expansion over radial angular variable, which limit accuracy and generate numerous spurious modes with small growth rates. Moreover, since the method uses an Eulerian formulation, it is subject to the same boundary‑integral requirement.

The aim of this paper is threefold. First, we show that the method can be generalized without increasing the size of the resulting matrix. Second, we assess the accuracy of the method by systematic convergence studies and comparison with an independent log-spiral method, 
and show that an accuracy of $\sim 0.003$ in both pattern speed and growth rate is achievable on modern GPU-equipped workstations.
Third, we make the implementation openly available on GitHub, together with several standard stellar‑disk models.

The paper is organized as follows. Section~2 reviews the linear matrix method and presents its generalization to distribution functions that are discontinuous at zero angular momentum (\emph{sharp-edge} DFs). Section~3 describes the Kuzmin--Toomre disk models used in this study, introduces taper functions that justify the generalized equation by demonstrating convergence in the sharp-edge limit, and outlines the numerical setup. Section~4 presents the results: taper convergence, extrapolation of eigenvalues over the harmonic truncation and grid parameters, and comparison with the log-spiral method. Section~5 discusses the results and summarizes the paper. Appendix~\ref{app:A} provides an alternative derivation of the generalized matrix equation, and Appendix~\ref{app:B} introduces the grid variables used in place of the conventional action variables.

\section{The linear  matrix method}

\subsection{The previous version}

In this subsection, we briefly review the derivation of the linear matrix equation presented in \cite{Pol05} and \cite{PJ15}.
The disk is razor-thin, so the phase space is four-dimensional with coordinates $(r,\theta,v_r,v_\theta)$, and the surface density is obtained by integrating the distribution function over the velocities $(v_r,v_\theta)$.
We write
\be
F(\br,\bbv)= F_0(\br,\bbv)+F_1(\br,\bbv;t),
\ee
\be\Phi(\br;t)=\Phi_0(r)+\Phi_1(\br;t),
\ee where $F$ is DF and $\Phi$ is gravitational potential, while $F_0$ and $\Phi_0$ are their equilibrium values.
The linearized collisionless Boltzmann equation connecting the perturbed DF $F_1$ with the perturbed potential $\Phi_1$ written in action-angle variables $(\bJ,\bw)\equiv (J_1,J_2; w_1,w_2)$, is 
\be
\frac{\p F_1}{\p t}+\Omega_1\,\frac{\p F_1}{\p w_1}+ \Omega_2\,\frac{\p F_1}{\p w_2}=\frac{\p\Phi_1}{\p w_1}\,\frac{\p F_0}{\p J_1}+
\frac{\p\Phi_1}{\p w_2}\,\frac{\p F_0}{\p J_2}.
\label{eq:CBE}
\ee
Here  
\begin{multline}
J_1=\frac{1}{\pi}\int\limits_{r_{\rm min}}^{r_{\rm max}}
v_r(r,E,L)\,\d r\\=\frac{1}{\pi}\int\limits_{r_{\rm min}}^{r_{\rm max}} \sqrt{2[E-\Phi_0(r)]-L^2/r^2}\,\d r
\label{eq:J1}
\end{multline}
is radial action, and
\be
J_2=L=r\,v_\theta
\ee
is azimuthal action, $w_1$ and $w_2$ are the corresponding conjugate angles (for their explicit form  see (\ref{eq:w1}) and (\ref{eq:w2})). 
In (\ref{eq:J1}) $E=\frac{1}{2}(v_r^2+v_\theta^2)+\Phi_0(r)$ and $L$ are the energy and angular momentum of the star in the unperturbed potential, respectively.  

Radial $\Omega_1$ and azimuthal $\Omega_2$ frequencies are $\Omega_{1,2}=\p E(J_1,J_2)/\p J_{1,2}$, 
which can be  found from (\ref{eq:J1}):
\be
\Omega_1(\bJ)=\frac{\pi}{\displaystyle\int_{r_{\rm min}}^{r_{\rm max}}\!\!\!\!{\d r}/{v_r}},\ \ 
\Omega_2(\bJ)=\frac{\Omega_1(\bJ)}{\pi}\int\limits_{r_{\rm min}}^{r_{\rm max}} \frac{L}{r^2\,v_r}\,\d r.
\ee
 Note that $J_1\ge 0$, but $J_2=L$ can have both signs. The negative $J_2$ corresponds to retrograde stars.  Also $\Omega_1$ is even, while $\Omega_2$ is odd function of $L$. In (\ref{eq:CBE}) $F_0$ is  expressed as a function of the integrals of motion  $J_1,J_2$. 
 
If we represent  the perturbed potential and DF as  $\Phi_1(r,\theta;t)\propto \exp(-\i\omega t+\i m\theta)$ and $F_1(r,\theta, v_r,v_\theta;t) \propto \exp(-\i\omega t+\i m\theta)$ and expand them  over angle variables
\be
\Phi_1=e^{-\i\omega t+\i m w_2}\,\sum\limits_{l=-\infty}^\infty \Phi_l(\bJ)\,e^{\i l w_1},
\label{eq:Phi_l}
\ee
\be
F_1=e^{-\i\omega t+\i m w_2}\,\sum\limits_{l=-\infty}^\infty \F_l(\bJ)\,e^{\i l w_1},
\label{eq:F_l}
\ee
we obtain the following equation connecting the amplitudes of harmonics $\F_l$ and $\Phi_l$:
\be
\Bigl[\omega-l\Omega_1(\bJ)-m\Omega_2(\bJ)\Bigr]\,\F_l(\bJ) = - F_{0,l}'(\bJ) \, \Phi_l(\bJ),
\label{eq:F_l1}
\ee
where
\be
 F_{0,l}'(\bJ) \equiv l\,\frac{\p F_0}{\p J_1}+ m\,\frac{\p F_0}{\p J_2} 
\label{eq:CBE_lin}
\ee
 In deriving (\ref{eq:F_l}) we used the fact that the angular variables $w_i$ are linearly dependent on time: $w_{1,2}=\Omega_{1,2}\,t+w_{1,2}^{(0)}$.

We use the Poisson equation in integral form
\be
\Phi_1=-G\int \d\br' \d\bbv' \frac{F_1(\br',\bbv')}{|\br-\br'|}
\label{eq:Phi1}
\ee
to express harmonics of the perturbed potential through the harmonics of the perturbed DF. This expression can be generalized to include gravitational softening, which is often used in N-body simulations. For Plummer softening, we simply replace $|\br-\br'|$ in the denominator of (\ref{eq:Phi1}) with
$\sqrt{(\br-\br')^2+\beta^2}$, where $\beta$ is the softening length.
The result is
\be
\Phi_l(\bJ)=-G\int \d\bJ'\sum\limits_{l'}\Pi_{l l'}(\bJ,\bJ')\,\F_{l'}(\bJ'),
\label{eq:Phil}
\ee
where
\begin{multline}
\Pi_{l l'}(\bJ,\bJ')=4\int\limits_0^\pi \d w_1 \int\limits_0^\pi \d w'_1
H_m\Bigl(r(w_1,\bJ), r(w_1',\bJ')\Bigr)\\
\times \cos(l w_1+m\phi)\,\cos(l' w_1'+m\phi').
\end{multline}
Here
\be H_m(r,r')=\frac{\psi_m(z)}{\pi\,\sqrt{r^2+r'^2+\beta^2}}\,
\ee
where 
\be
\psi_m(z)=\int\limits_0^\pi \frac{\cos(m\alpha)\,\d\alpha}{\sqrt{1-z\,\cos\alpha}},\ \ z=\frac{2\,rr'}{r^2+r'^2+\beta^2}
\ee

The function $\psi_m(z)$ can be expressed through the complete elliptic integrals of the first and the second kind, ${\bf K}(\kappa)$ and  ${\bf E}(\kappa)$ respectively, \citep[see][]{Gradshteyn_8}. In particular, for $m=2$
\begin{multline}
\psi_2(z)=\frac{2}{\sqrt{1+z}}\left[\left(1-\frac{16}{3\kappa^2}+\frac{16}{3\kappa^4}\right){\bf K}(\kappa)\right.\\ \left.+\left(\frac{8}{3\kappa^2}-\frac{16}{3\kappa^4}\right){\bf E}(\kappa)\right],
\end{multline}
where \be
\kappa^2=\frac{2z}{1+z}=\frac{4rr'}{(r+r')^2+\beta^2}.
\ee

The dependence $r(w_1,\bJ)$ is defined by implicit relation 
\be
w_1(r,\bJ)=\Omega_1 \int\limits_{r_{\rm min}(\bJ)}^r\frac{\d r'}{ v_r(r',\bJ)},
\label{eq:w1}
\ee
while $w_2$ is connected with polar angle $\theta$ of the star as
\be 
w_2=\theta+\phi(w_1,\bJ),
\label{eq:w2}
\ee
where
\be
\phi(w_1,\bJ)=\frac{\Omega_2}{\Omega_1}\,w_1-J_2\!\!\!\!\int\limits_{r_{\rm min}(\bJ)}^{r(w_1,\bJ)}\frac{\d r'}{r'^2 v_r(r',\bJ)}.
\label{eq:phi}
\ee
Substitution of (\ref{eq:Phil}) into (\ref{eq:F_l}) yields
\begin{multline}
    \Bigl[\omega-l\Omega_1(J,L)-m\Omega_2(J,L)\Bigr]\,{\cal F}_l(J,L)\\
    = F_{0,l}'(\bJ) \int dJ' dL'\sum\limits_{l'}\Pi_{ll'}(J,L;J',L')\,{\cal F}_{l'}(J',L').
    \label{eq:Zh2}
    \end{multline}
Hereafter we denote the radial action $J_1$ as $J$ and the azimuthal action $J_2$ as $L$.

In order to reduce (\ref{eq:Zh2}) to the matrix equation
\be 
\omega\, \textsf{x = A\,x}
\label{eq:LME}
\ee
one must specify the grid in action sub-space $(J,L)$, or in equivalent 2D sub-space, chosen  for convenience, say, $(E,L)$, or $(R_c,\cc)$ (for $R_c$ and $\cc$ variables see App.\,\ref{app:B}), and the cutoff parameters
($\ell_{\rm min},\,  \ell_{\rm max}$) for the Fourier expansions (\ref{eq:Phi_l}) and (\ref{eq:F_l}). To get accurate results, one needs to take a sufficiently large range of summation on $l$ into account. Their total number $N_l = \ell_{\rm max} - \ell_{\rm min} + 1$ depends on the model under investigation. The order of the matrix $\textsf{A}$ equals $N_l\times N_{\rm ph}$, where $N_{\rm ph}$ is the number of mesh points in the chosen 2D phase plane.

Unfortunately, the matrix equation derived from the system of integral equations (\ref{eq:Zh2}) is inapplicable in the case where the unperturbed DF $F_0(J,L)$  has a sharp edge on radial orbits, $F_0(J,L)={\mathfrak F}(J,L)\Theta(L)$, where $\Theta(z)$ is the Heaviside step function. Indeed, in such a case the derivative $\p F_0/\p L$ in r.h.s. of  (\ref{eq:Zh2}) contains $\delta(L)$, which cannot be correctly included in the matrix equation (\ref{eq:LME}). A modification of it is required.

\subsection{Generalization}
    
The simplest generalization of (\ref{eq:Zh2}) to an unperturbed DF with a sharp edge at radial orbits, $L=0$, is to make the substitution in (\ref{eq:Zh2}):
\be
    h_l(J,L)= \frac{{\cal F}_l(J,L)}{l\,{\p F_0}/{\p J}+m\,{\p F_0}/{\p L}}.
\ee
Then we obtain
\begin{multline}
\Bigl[\omega-l\Omega_1(J,L)-m\Omega_2(J,L)\Bigr]\,h_l(J,L)\\=\!\!
\int dJ'\int dL'\sum\limits_{l'}\Pi_{l l'}(J,L;J'\!\!,L')\\
\times \Bigl(l'\,\frac{\p F_0}{\p J'}+m\,\frac{\p F_0}{\p L'}\Bigr)\,h_{l'}(J'\!\!,L').
\label{eq:h_l}
\end{multline}
For the DF  with sharp edge at radial orbits $L=0$, 
\be
F_0(J,L)={\mathfrak F}(J,L)\,\Theta(L),
\ee 
we have
\begin{multline}
l\,\frac{\p F_0}{\p J}+m\,\frac{\p F_0}{\p L}=
\left[l\,\dfrac{\p {\mathfrak F}(J,L)}{\p J}+m\,\dfrac{\p {\mathfrak F}(J,L)}{\p L}\right]\,\Theta(L)\\+m\,{\mathfrak F}(J,0)\,\delta(L),
\label{eq:F_0l}
\end{multline}
and finally obtain
\begin{multline} 
\Bigl[\omega-l\Omega_1(J,L)-m\Omega_2(J,L)\Bigr]\,h_l(J,L)\\= 
\int dJ'\Biggl[\,\,\int\limits_{L'>0} dL'\sum\limits_{l'}\Pi_{l l'}(J,L;J',L')\,\\ \times \Bigl(l'\,\frac{\p {\mathfrak F}}{\p J'}+m\,\frac{\p {\mathfrak F}}{\p L'}\Bigr)\,h_{l'}(J',L')\\+\Pi_{ll'}(J,L;J',0)\, {\mathfrak F}(J',0)\,m\,h_{l'}(J',0)\Biggr].
\label{eq:final} 
\end{multline} 
Eq. (\ref{eq:final}) is the generalization of (\ref{eq:Zh2})  for the case of a DF with a sharp edge at  $L=0$, i.e, ${\mathfrak F}(J,L=0)\ne 0$. The sharp edge at radial orbits leads to the appearance of an additional term on the r.h.s. of (\ref{eq:final}). Now this equation can be easily converted to matrix form (\ref{eq:LME}).  The alternative and less formal derivation of the modified Eq. (\ref{eq:final}) is given in Appendix \ref{app:A}. 

\section{Models and numerical setup}

\subsection{Kuzmin--Toomre disk models}
\label{subsec:KT_models}

The gravitational potential $\Phi_0$ and surface density $\sigma_0$ of the Kuzmin--Toomre disk are given by \cite{Kuzmin56} and \cite{Toomre63}:
\be
\Phi_0(r)=-\frac{GM}{\sqrt{b^2+r^2}},\ \  \sigma_0(r)=\frac{M\,b}{2\pi\,(b^2+r^2)^{3/2}}.
\ee
We consider two unidirectional DF families for this potential, suggested by \cite{Miyamoto71} and \cite{Kalnajs76}. In what follows, we set $M=b=G=1$.

\subsubsection{The Miyamoto/Hunter family}

\cite{Miyamoto71} constructed a model that includes retrograde stars and expressed the DF as the sum of functions symmetric and antisymmetric in $L$. \cite{Hunter92} retained only the symmetric part for $L>0$, doubling the normalization constant. For comparison, we adopt the form given in \cite{Hunter92}. This is a one-parameter family of DFs, labeled by an integer $n_{\rm M}\ge 1$:
\begin{multline}
F_{\rm M}(E,L)={\mathfrak F}_{\rm M}(E,L)\,\Theta(L)\\ \equiv\, \frac{n_{\rm M}!\,\Gamma(2\,n_{\rm M}+4)}{2\pi^2}\,(-E)^{2n_{\rm M}+2} \times \\
\sum\limits_{s=0}^{n_{\rm M}}\frac{2^s} {s!\,(n_{\rm M}-s)!\,(2s-1)!!\, \Gamma(2\,n_{\rm M}-s+3)}\,\left(\frac{L^2}{-2E}\right)^s\!\!\Theta(L).    
\end{multline}
It can be written in a more compact form \citep[][]{Hunter92}
\begin{multline}
  F_{\rm M}(E,L)={\mathfrak F}_{\rm M}(E,L)\,\Theta(L)\\ \equiv\frac{2n_{\rm M}+3}{2\pi^2}(-E)^{2n_{\rm M}+2} 
  F\Bigl(\!-n_{\rm M},\!-2n_{\rm M}-2,\frac{1}{2};\!\frac{L^2}{-2E}\Bigr)\Theta(L),
  \nonumber
\end{multline}
where  $F(a,b,c;z)$  is a hypergeometric function~\citep[see, e.g.][]{Gradshteyn_8}.

\subsubsection{The Kalnajs family}

The  \cite{Kalnajs76} family is also a  one-parameter family of DFs, parameterized by an integer $m_K\ge 1$, which regulates the dynamical temperature: the larger $m_K$, the colder the disk, i.e.\ the smaller the radial velocity dispersion $\langle v_r^2\rangle^{1/2}$:
\be
F_{\rm K}(E,L)={\mathfrak F}_{\rm K}(E,L)\,\Theta(L)\equiv (-2 E)^{m_{\rm K}-1} g_{m_{\rm K}}(x)\,\Theta(L),
\ee
where $x=\sqrt{-2E L^2}$ and
\begin{multline}
    g_{m_{\rm K}}(x) =\frac{1}{\pi} \left[x\,\frac{d\tau_{m_{\rm K}}}{dx}-
\frac{m_{\rm K} (m_{\rm K}-3)}{2}\,\tau_{m_{\rm K}}(x) \right. 
+ \\
\left. \int_0^1\!\!\!\d\eta\,\tau_{m_{\rm K}}(\eta\,x)\,\eta^{m_{\rm K}}\, P^{''}_{m_{\rm K}-1}(\eta)\right], \ \  g_{m_{\rm K}}(0)=\dfrac{m_{\rm K}}{2^{m_{\rm K}}\pi^2},
\label{eq:g_mK}
\end{multline}
where $P_n^{''}(z)$ is the second derivative of the Legendre polynomial and
$\tau_{m_{\rm K}}(x)$ is
\be
\tau_{m_{\rm K}}(x)=\frac{1}{2\pi}\,(1-x^2)^{(3-m_{\rm K})/2}.
\ee

In the general case of potential-density pairs with a softened central potential, $|\Phi_0(0)|={\rm const}\ne \infty$, the function $\tau_{m_{\rm K}}(y)$ is obtained from
\be
    \sigma_0(r)=w^{m_{\rm K}}\tau_{m_{\rm K}}(y)\,,
\ee
where
\be
    w=\frac{\Phi_0(r)}{\Phi_0(0)},\quad
    y=\frac{r\Phi_0(r)}{r_{\ast}\Phi_0(0)},\quad
    r_{\ast}=\lim\limits_{r\to \infty} \frac{r\,\Phi_0(r)}{\Phi_0(0)}.
\ee

These models, together with the \cite{Kalnajs76} and \cite{PLB96} families for the isochrone potential \citep[see][]{Henon59, BT2008}, are among the most widely studied and have been used to compute normal modes via both eigenvalue methods and $N$-body simulations by several authors \citep[see e.g.,][]{Kalnajs78, Z&H78, Hunter92, E&S95, AS86, PC97, Pol05, JH05, Pol_13, PJ15, RFD19}.

We consider bisymmetric ($m=2$) normal modes. The selected parameters, $n_{\rm M}=3$ for the Miyamoto family and $m_{\rm K}=6$ for the Kalnajs family, guarantee stability against axisymmetric disturbances: as shown by \cite{Miyamoto71} and \cite{Kalnajs76}, for these parameters the Toomre stability criterion is satisfied, $Q>1$.

\subsection{Tapered DF family and the sharp-edge limit}
\label{subsec:tapers}

To avoid the abrupt discontinuity of the DF at $L=0$, one commonly applies so-called tapers that reverse the orbits of a fraction of stars with small angular momenta, keeping the surface density unchanged and smoothing the discontinuity \citep{Z&H78, AS86, SA86, Pol_13}. These authors use a cubic taper in $L$ with a characteristic scale $L_\ast$. However, this taper is inconvenient for studying the sharp-edge limit: the angular momentum ranges from zero near the centre to values of order unity at large radii, so any fixed $L_\ast$ is too large for stars near the centre and too coarse for the grid resolution farther out.

Hereafter we work in the phase space $(R_c,\cc)$, where $R_c = (r_{\min}+r_{\max})/2$, analogous to the semi-major axis of the Keplerian ellipse, and the dimensionless variable $\cc$ is the {\it circularity} -- a quantity complementary to eccentricity (see Appendix~\ref{app:B} for details). It varies in the range $-1\le\cc\le 1$, where $|\cc|=1$ corresponds to circular orbits and $\cc=0$ to radial ones. Since $\cc$ is uniformly bounded, a taper defined in terms of $\cc$ avoids the scale problem. It takes the form
\be
     T_{\cc_\ast}(\cc)=\left\{
    \ba{ll}
    0,& \cc<-\cc_{\ast},\\
\frac{1}{4}\bigl[2+3\,(\cc/\cc_{\ast})-({\cc}/{\cc_{\ast}})^3\,\bigr],&|\cc|<\cc_{\ast},\\
1,& \cc>\cc_{\ast},
\ea
\right.
\label{eq:poly}
\ee
and the smoothed DF is $\tilde{F}_{\cc_\ast} = \mathfrak{F}(E,L)\,T_{\cc_\ast}(\cc)$. Note that $\mathfrak{F}_{\rm M}$ and $\mathfrak{F}_{\rm K}$, considered as functions of $R_c$ and $\cc$, are even functions of $\cc$. This taper reverses the motion of all stars with circularity $\cc < \cc_\ast$.

For any finite $\cc_\ast > 0$, the smoothed DF is continuous and the non-generalized matrix equation~(\ref{eq:Zh2}) applies without modification. In the limit $\cc_\ast \to 0$, the taper approaches the Heaviside step function, $T_{\cc_\ast}(\cc) \to \Theta(\cc)$, and $\tilde{F}_{\cc_\ast} \to F_0 = \mathfrak{F}(E,L)\,\Theta(L)$. By computing eigenvalues for a sequence of decreasing $\cc_\ast$ and verifying their convergence to the values obtained from the generalized equation~(\ref{eq:final}), we validate the generalization (Section~\ref{sec:results}).

\subsection{Ambiguity of $\Omega_2$ at radial orbits}
\label{subsec:omega2}

The generalized equation~(\ref{eq:final}) contains the boundary term evaluated at $L=0$, where the azimuthal frequency $\Omega_2(J,0)$ is discontinuous: $\Omega_2(J,0)=\frac{1}{2}\,\Omega_1(J,0)\,{\rm sgn}(L)$. This means that the frequency $\Omega_2(J,0)$ appearing on the l.h.s.\ of~(\ref{eq:final}) is ambiguous when $L=0$. The taper convergence described above resolves this ambiguity: as $\cc_\ast \to 0$, the eigenvalues converge to the values obtained from~(\ref{eq:final}) with
\be
\Omega_2(J,0)=\lim\limits_{L\to +0} \Omega_2(J,L)=\case{1}{2}\,\Omega_1(J,0)>0.
\ee

However, a numerical experiment with $\Omega_2(J,0)=-\frac{1}{2}\,\Omega_1(J,0)$ in~(\ref{eq:final}) gives the same eigenvalues, to within the numerical accuracy, as $\Omega_2(J,0)=\frac{1}{2}\,\Omega_1(J,0)$, so the ambiguity does not affect the eigenvalues. It can be rigorously shown that both choices leave the system (\ref{eq:final}) unchanged for $\ell_{\rm max}=\infty$ and, for finite $\ell_{\rm max}$, differ only by a shift of the summation limits in $l$, which is negligible when $\ell_{\rm max}$ is large enough. For even $m$ the same holds also for the substitution $\Omega_2(J,0)=0$ in  (\ref{eq:final}).

The same ambiguity problem also exists in the classical \cite{Kalnajs77} matrix equation when applied to unidirectional DFs with a sharp edge at $L=0$, regardless of whether it is written in its original {\it Lagrangian} form or transformed into {\it Eulerian} form \citep[as in][]{JH05}.
The proven independence of the eigenvalues from the choice of  $\Omega_2(J,0)$ solves the ambiguity problem also  for the application of the \cite{Kalnajs77} matrix equation to sharp-edge models.

\subsection{Numerical setup}
\label{subsec:setup}

The phase-space grid is constructed in the $(R_c,\cc)$ plane (see Appendix~\ref{app:B}). The radial grid consists of $N_R$ points equally spaced in $u=\ln R_c$ over the interval $[R_{\rm min},\,R_{\rm max}]$. The circularity grid depends on the model type. For sharp-edge DFs, $N_\cc$ points are uniformly spaced in $\cc\in[0,1]$ with Simpson's-rule quadrature weights. For tapered DFs, the grid is split into two segments: $N_{\cc,t}$ points uniformly spaced in $[-\cc_\ast,\,\cc_\ast]$ (covering the taper transition region) and $N_\cc - N_{\cc,t}+1$ points in $[\cc_\ast,\,1]$, with Simpson's rule applied independently to each segment and weights summed at the junction point $\cc_\ast$.

Orbital frequencies $\Omega_1$, $\Omega_2$, the radial action $J$, and the trajectory $r(w_1)$ are computed by numerical integration along the orbit using $N_{w,{\rm orbit}}$ phase points; these quantities are then sampled on a coarser grid of $N_w$ points for the assembly of the matrix elements $\Pi_{ll'}$.

The perturbed DF and potential are expanded in Fourier harmonics with indices $l$ running from $\ell_{\rm min}=-\ell_{\rm max}$ to $\ell_{\rm max}$, giving $N_l=2\ell_{\rm max}+1$ harmonics. The resulting matrix $\textsf{A}$ in~(\ref{eq:LME}) has order $N_l\times N_R\times N_\cc$. Gravitational interactions are softened with the Plummer parameter $\beta$.

For sharp-edge calculations, the standard parameter values are $N_R=61$, $N_\cc=21$, $N_w=501$, $N_{w,{\rm orbit}}=10^5$, $\beta=10^{-5}$, $R_{\rm min}=0.01$, $R_{\rm max}=20$, and $\ell_{\rm max}=7$ (unless stated otherwise). For tapered calculations, the circularity grid uses $N_\cc=41$ total points with $N_{\cc,t}=21$ in the taper region. The eigenvalues of $\textsf{A}$ are computed via full LAPACK decomposition. The implementation uses Julia with GPU-accelerated computation of the matrix elements $\Pi_{ll'}$.

\section{Results}
\label{sec:results}

\subsection{Taper convergence to the sharp-edge limit}

\begin{figure*}
\centering
\includegraphics[width=86mm]{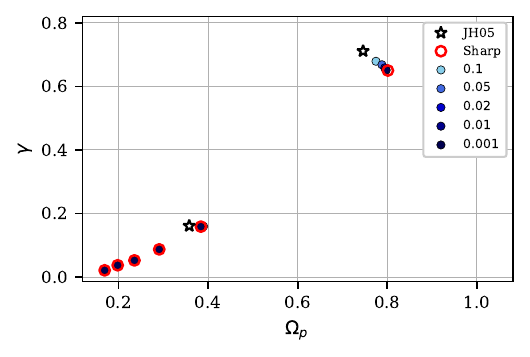}\hspace{5mm}
\includegraphics[width=86mm]{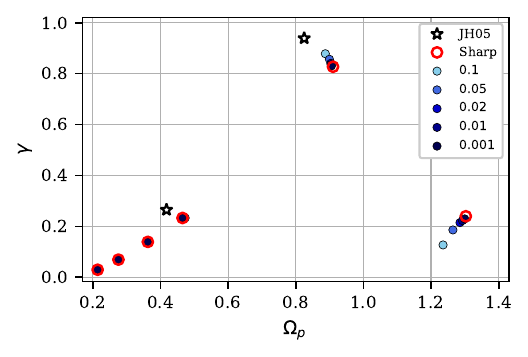}
\caption{Pattern speed $\Omega_p$ versus growth rate $\gamma$ for the Kalnajs (1976) $m_{\rm K}=6$ model (left panel) and for the Miyamoto (1971) $n_M=3$ model (right panel) of the Kuzmin-Toomre disk. Black stars show the JH05 reference values. Red circles represent sharp boundary calculations (61$\times$21$\times$7). Colored circles show polynomial taper results for five different minimum circularity values: $\cc_\ast = 0.1, 0.05, 0.02, 0.01, 0.001$ (61$\times$41$\times$7), with colors progressing from light blue (high $\cc_\ast$) to dark blue (low $\cc_\ast$).}
\label{fig:kuzmin_mk6_miyamoto_nm3}
\end{figure*}

Tables~\ref{tab:kalnajs_modes} and \ref{tab:miyamoto_modes} list the five most unstable eigenvalues $(\Omega_p, \gamma)$ for the Kalnajs $m_{\rm K}=6$ and Miyamoto $n_{\rm M}=3$ models; Figure~\ref{fig:kuzmin_mk6_miyamoto_nm3} shows six modes for each model, including a sixth mode not tabulated. As $\cc_\ast \to 0$, the polynomial-taper eigenvalues converge to the sharp-edge values obtained from the generalized equation~(\ref{eq:final}) with $\Omega_2(J,0)=\frac{1}{2}\,\Omega_1(J,0)$, confirming its validity.

\begin{table*}
\setlength{\tabcolsep}{3pt}
\begin{tabular*}{\textwidth}{@{\extracolsep{\fill}}lccccccc}
\hline\hline
Configuration & Grid ($N_R\!\times\!N_\cc\!\times\!\ell_{\rm max}$) & Mode 1 & Mode 2 & Mode 3 & Mode 4 & Mode 5 \\
\hline
Poly $\cc_\ast=0.1$    & 61$\times$41$\times$7 & (0.775, 0.679) & (0.390, 0.160) & (0.292, 0.089) & (0.236, 0.052) & (0.199, 0.037) \\
Poly $\cc_\ast=0.05$   & 61$\times$41$\times$7 & (0.788, 0.668) & (0.387, 0.159) & (0.291, 0.088) & (0.236, 0.052) & (0.198, 0.037) \\
Poly $\cc_\ast=0.02$   & 61$\times$41$\times$7 & (0.794, 0.658) & (0.384, 0.159) & (0.291, 0.087) & (0.236, 0.052) & (0.198, 0.037) \\
Poly $\cc_\ast=0.01$   & 61$\times$41$\times$7 & (0.796, 0.655) & (0.384, 0.159) & (0.291, 0.087) & (0.236, 0.052) & (0.198, 0.037) \\
Poly $\cc_\ast=0.001$  & 61$\times$41$\times$7 & (0.798, 0.651) & (0.383, 0.159) & (0.291, 0.087) & (0.235, 0.052) & (0.198, 0.037) \\[1mm]
Sharp             & 61$\times$21$\times$7 & (0.801, 0.650) & (0.384, 0.159) & (0.291, 0.087) & (0.236, 0.052) & (0.198, 0.037) \\
Sharp             & 61$\times$21$\times$15 & (0.814, 0.655) & (0.389, 0.161) & (0.294, 0.089) & (0.238, 0.053) & (0.200, 0.037) \\
Sharp             & 61$\times$21$\times\infty$ & (0.817, 0.657) & (0.390, 0.162) & (0.295, 0.089) & (0.238, 0.053) & (0.200, 0.037) \\[1mm]
Exact             & -- & (0.813, 0.653) & (0.390, 0.161) & (0.295, 0.090) & (0.239, 0.050) & (0.202, 0.033) \\[1mm]
JH05              & -- & (0.746, 0.711) & (0.358, 0.161) & -- & -- & -- \\
Sharp (no boundary term)     & 61$\times$21$\times$15 & (0.339, 0.298) & (0.341, 0.250) & (0.270, 0.180) & (0.288, 0.164) & (0.223, 0.159) \\
\hline
\end{tabular*}
\caption{Eigenvalues $(\Omega_p, \gamma)$ for the Kalnajs $m_{\rm K}=6$ model of the Kuzmin-Toomre disk. Grid notation: $N_R$ = radial points; $N_\cc$ = circularity points; $|\ell_{\rm min}|=\ell_{\rm max}$ = maximum angular harmonic index (total number of angular harmonics is $N_l=2\ell_{\rm max}+1$). `Sharp (wrong)' means incorrect eigenvalues calculated  without taking into account $\delta(L)$ in (\ref{eq:F_0l}).}
\label{tab:kalnajs_modes}
\end{table*}

\begin{table*}
\setlength{\tabcolsep}{3pt}
\begin{tabular*}{\textwidth}{@{\extracolsep{\fill}}lccccccc}
\hline\hline
Configuration & Grid ($N_R\!\times\!N_\cc\!\times\!\ell_{\rm max}$) & Mode 1 & Mode 2 & Mode 3 & Mode 4 & Mode 5  \\
\hline
Poly $\cc_\ast=0.1$    & 61$\times$41$\times$7 & (0.888, 0.878) & (0.473, 0.233) & (0.364, 0.145) & (0.277, 0.070) & (1.236, 0.127) \\
Poly $\cc_\ast=0.05$   & 61$\times$41$\times$7 & (0.900, 0.856) & (0.469, 0.233) & (0.363, 0.142) & (0.276, 0.069) & (1.265, 0.186) \\
Poly $\cc_\ast=0.02$   & 61$\times$41$\times$7 & (0.904, 0.841) & (0.466, 0.233) & (0.363, 0.140) & (0.276, 0.069) & (1.286, 0.215) \\
Poly $\cc_\ast=0.01$   & 61$\times$41$\times$7 & (0.905, 0.835) & (0.465, 0.234) & (0.362, 0.139) & (0.276, 0.069) & (1.293, 0.223) \\
Poly $\cc_\ast=0.001$  & 61$\times$41$\times$7 & (0.906, 0.830) & (0.465, 0.234) & (0.362, 0.139) & (0.275, 0.069) & (1.300, 0.229) \\[1mm]
Sharp             & 61$\times$21$\times$7 & (0.910, 0.827) & (0.465, 0.233) & (0.363, 0.139) & (0.276, 0.069) & (1.303, 0.240) \\
Sharp             & 61$\times$21$\times$15 & (0.921, 0.820) & (0.470, 0.239) & (0.369, 0.146) & (0.281, 0.073) & (1.317, 0.295) \\
Sharp             & 61$\times$21$\times\infty$ & (0.924, 0.819) & (0.472, 0.240) & (0.370, 0.148) & (0.283, 0.074) & (1.320, 0.310) \\[1mm]
Exact             & -- & (0.915, 0.817) & (0.470, 0.239) & (0.370, 0.147) & (0.283, 0.074) & (1.325, 0.298) \\[1mm]
JH05              & -- & (0.825, 0.939) & (0.418, 0.265) & -- & -- & -- \\
Sharp (no boundary term)     & 61$\times$21$\times$15 & (0.357, 0.293) & (0.294, 0.101) & (0.217, 0.043) & (0.171, 0.021) & (0.146, 0.011) \\
\hline
\end{tabular*}
\caption{The same as in Table\,\ref{tab:kalnajs_modes}, for the Miyamoto/Hunter $n_M=3$ model of the Kuzmin-Toomre disk}
\label{tab:miyamoto_modes}
\end{table*}

For the Kalnajs model, the dominant mode (mode~1) is the most sensitive to $\cc_\ast$, while modes~2--5 are practically converged already at $\cc_\ast=0.1$. The Miyamoto model shows a similar pattern, except that both mode~1 and mode~5 are sensitive to $\cc_\ast$. Mode~5, with pattern speed $\Omega_p>1$ exceeding the maximum circular frequency in the disk, is unusual but not a numerical artifact: it persists across all grid resolutions and taper values.

Figure~\ref{fig:kuzmin_mk6_miyamoto_nm3} summarizes these results in the $(\Omega_p,\gamma)$ plane. For each mode, the colored circles trace the taper convergence from $\cc_\ast=0.1$ (light blue) to $\cc_\ast=0.001$ (dark blue), approaching the sharp-edge value (red circle). In the Kalnajs model (left panel), the taper sequence for modes~2--5 is so tightly clustered that the individual circles overlap, whereas mode~1 shows a clear drift toward the sharp-edge value as $\cc_\ast$ decreases. In the Miyamoto model (right panel), mode~1 behaves similarly, and mode~5 forms a separate cluster at high $\Omega_p$. The JH05 values (black stars) are visibly displaced from both the taper sequence and the sharp-edge values, particularly for mode~1.

To assess the accuracy of the linear method, we also compute eigenvalues using an independent approach that expands the perturbed surface density and potential in logarithmic spirals~\citep[see e.g.][]{BT2008}, leading to a nonlinear eigenvalue problem in $\omega$. This approach is similar to that of \cite{Zang76}, generalized here to arbitrary potentials; we refer to it as the \emph{log-spiral method}. It converges much faster than the linear method, and it appears feasible to achieve \texttt{x.xxxx} accuracy or better (we quote accuracies using a fixed-precision convention: \texttt{x.xxx} and \texttt{x.xxxx} denote results that are firm to three and four decimal places, respectively); a detailed description will be given in a separate paper. The rows labeled `Exact' in Tables~\ref{tab:kalnajs_modes} and \ref{tab:miyamoto_modes} give the eigenvalues obtained by this method. For both models, the sharp-edge values extrapolated to $\ell_{\rm max}\to\infty$ agree with the exact values to within ${\sim}\,0.005$ for most modes.

The rows labeled `Sharp (wrong)' demonstrate the necessity of the generalization: omitting the boundary term $\delta(L)$ in~(\ref{eq:F_0l}) produces eigenvalues that bear no resemblance to the correct ones.

The \cite{JH05} values differ substantially from both our results and the exact values, especially for mode~1 of the Miyamoto model ($\Omega_p=0.825$ vs $0.915$, $\gamma=0.939$ vs $0.817$). According to M.A.~Jalali (private communication), their calculations aimed at ${\sim}\,10\%$ accuracy. Our results show that significantly higher accuracy is attainable for all modes.

\subsection{Convergence and extrapolation of eigenvalues}
\label{subsec:convergence}

The eigenvalues computed by the linear matrix method depend on several numerical parameters: the harmonic truncation $\ell_{\rm max}$, the number of radial grid points $N_R$, the grid spacing $\Delta u$ in the variable $u = \ln R_c$, and the grid boundaries $R_{\rm min}$, $R_{\rm max}$, as well as quadrature parameters ($N_w$, $N_{w,{\rm orbit}}$) and the softening length $\beta$. We adopt the following strategy: fix the quadrature and softening parameters at values sufficiently large (or small) that they do not affect the eigenvalues to \texttt{x.xxxx}  accuracy, and then study the remaining corrections -- due to $\ell_{\rm max}$, $N_R$, and $\Delta u$ -- assuming that they are mutually independent.

\begin{table}[t]
\label{tab:convergence}
\setlength{\tabcolsep}{4pt}
\begin{tabular}{lcccc}
\hline\hline
$\ell_{\rm max}$ & \multicolumn{2}{c}{Kalnajs $m_{\rm K}=6$} & \multicolumn{2}{c}{Miyamoto $n_M=3$} \\
\cline{2-3} \cline{4-5}
 & $\Omega_p$ & $\gamma$ & $\Omega_p$ & $\gamma$ \\
\hline
3  & 0.773 & 0.597 & 0.874 & 0.785 \\
5  & 0.791 & 0.637 & 0.898 & 0.822 \\
7  & 0.801 & 0.650 & 0.910 & 0.827 \\
9  & 0.807 & 0.654 & 0.916 & 0.826 \\
11 & 0.811 & 0.655 & 0.919 & 0.824 \\
13 & 0.813 & 0.655 & 0.920 & 0.822 \\
15 & 0.814 & 0.655 & 0.921 & 0.820 \\
$\infty$ & 0.817 & 0.657 & 0.924 & 0.819 \\
\hline
\end{tabular}
\caption{Convergence of the dominant mode eigenvalue $(\Omega_p, \gamma)$ as a function of angular harmonic truncation $\ell_{\rm max}$ (with $\ell_{\rm min}=-\ell_{\rm max}$). Grid: $61 \times 21$.}
\end{table}

\begin{figure}[t]
\centering
\includegraphics[width=\columnwidth]{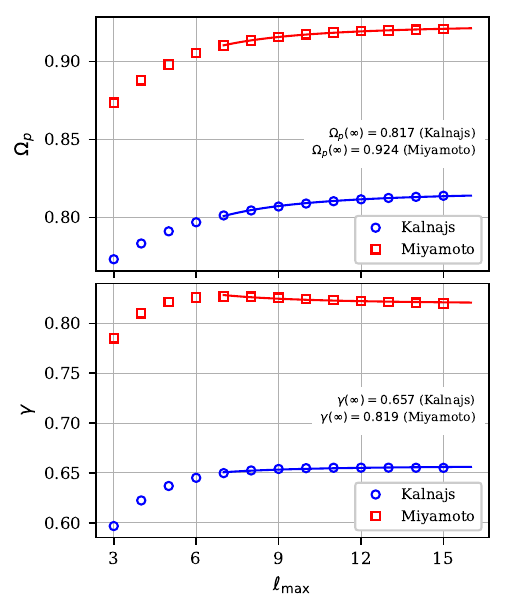}
\caption{Convergence of the dominant mode eigenvalue with angular harmonic truncation $\ell_{\rm max}$ for the Kalnajs $m_{\rm K}=6$ (blue circles) and Miyamoto $n_M=3$ (red squares) Kuzmin-Toomre disk models. Upper panel: pattern speed $\Omega_p$; lower panel: growth rate $\gamma$. Solid lines show power-law fits $a + b/\ell_{\rm max}^2$ to data with $\ell_{\rm max} \geq 7$, yielding asymptotic values $\Omega_p(\infty)$ and $\gamma(\infty)$. Grid: $61 \times 21$, sharp boundary.}
\label{fig:kres_convergence}
\end{figure}

Figure~\ref{fig:kres_convergence} shows the convergence of the dominant
mode eigenvalue as a function of the angular harmonic truncation
$\ell_{\rm max}$. Both the pattern speed $\Omega_p$ and growth rate
$\gamma$ exhibit systematic convergence with increasing resolution.
We fit the data to a power law $a + b/\ell_{\rm max}^2$, assuming
second-order convergence, to extrapolate the asymptotic eigenvalues.
The extrapolated values are $(\Omega_p, \gamma) = (0.817, 0.657)$ for
the Kalnajs $m_{\rm K}=6$ model and $(0.924, 0.819)$ for the Miyamoto $n_{\rm M}=3$ model.

We now turn to the grid-parameter corrections. The radial grid points $(R_c)_i$ are evenly spaced in $u = \ln R_c$ with step $\Delta u$. For each value of $\Delta u$, we extrapolate the eigenvalues to $N_R \to \infty$, and then extrapolate the resulting limiting values to $\Delta u \to 0$. Because this procedure is computationally expensive, we carry it out only for the dominant mode of the Miyamoto $n_{\rm M}=3$ model. The pattern speed exhibits a linear dependence $a + b\,\Delta u$ in the range $\Delta u = 0.12$--$0.06$; for smaller $\Delta u$, the required $N_R$ exceeds computational limits. The growth rate converges faster and becomes constant within \texttt{x.xxx} accuracy already at $\Delta u \leq 0.08$. The extrapolated values at $\Delta u = 0$ are $(\Omega_p, \gamma) = (0.904, 0.828)$.
Combining the $\ell_{\rm max}$ correction (computed on the $61 \times 21$ grid with $R_{\rm min} = 0.01$, $R_{\rm max} = 20$, yielding $+0.014$ for $\Omega_p$ and $-0.008$ for $\gamma$) with the grid-parameter extrapolation yields a final estimate of $(\Omega_p, \gamma) = (0.918, 0.820)$. For the same mode, the log-spiral method yields $(\Omega_p, \gamma) = (0.915, 0.817)$, so the two results differ by approximately $0.003$ in both components. To our knowledge, this is the first systematic error estimate for eigenvalues computed by the matrix method.

\begin{figure}[t]
\centering
\includegraphics[width=\columnwidth]{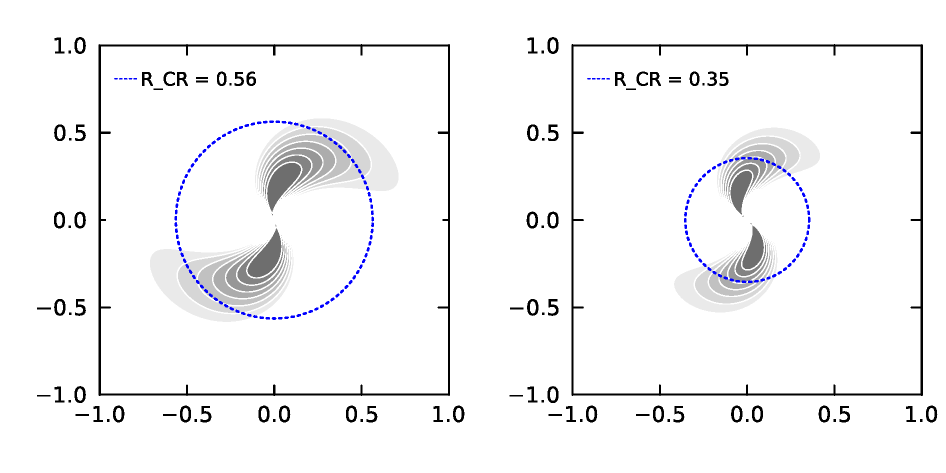}
\caption{Eigenfunctions of the most unstable $m=2$ mode. \emph{Left:} the Kalnajs $m_{\rm K}=6$ model. \emph{Right:} the Miyamoto $n_{\rm M}=3$ model. Colours show the real part of the perturbed surface density $\Sigma_1(r,\theta)$, normalised to its peak value.}
\label{fig:eigenfunctions}
\end{figure}

Figure~\ref{fig:eigenfunctions} shows the eigenfunctions of the most unstable $m=2$ mode for the two models: the Kalnajs $m_{\rm K}=6$ model in the left panel and the Miyamoto $n_{\rm M}=3$ model in the right panel. The colour scale gives the real part of the perturbed surface density $\Sigma_1(r,\theta)$ in the disc plane at a fixed instant, normalised to its peak value. The pattern rotates with the angular speed $\Omega_p$ and grows at the rate $\gamma$ listed for mode~1 in Tables~\ref{tab:kalnajs_modes} and \ref{tab:miyamoto_modes} (`Exact'). In both models the most unstable mode has the form of an open trailing spiral, which tightens into a bar during the subsequent nonlinear evolution.

\section{discussion and summary}
\label{sec:conclusions}

We have generalized the linear matrix method of \citet{Pol05} to distribution functions with sharp edges at zero angular momentum. The generalization introduces boundary-integral terms arising from the $\delta(L)$ contribution to the derivative of the DF, without increasing the size of the matrix. An alternative derivation of the generalized equation, based on explicit perturbation of the phase-space boundary, is given in Appendix~\ref{app:A}. The validity of the generalized equation
is confirmed by the smooth convergence of tapered-DF eigenvalues to the sharp-edge values as the taper width $\cc_\ast\to 0$.

We have applied the method to two Kuzmin--Toomre disk models: the Miyamoto $n_{\rm M}=3$ and Kalnajs $m_{\rm K}=6$ families. At least six unstable $m=2$ modes are found for each model (five are tabulated, a sixth is visible in Figure~\ref{fig:kuzmin_mk6_miyamoto_nm3}); the sequence of modes with decreasing pattern speed and growth rate likely continues, and additional modes outside this sequence may exist, as illustrated by mode~5 of the Miyamoto model. A systematic convergence study over the harmonic truncation $\ell_{\rm max}$ and the radial grid parameters $(\Delta u, N_R)$ yields, for the dominant mode of the Miyamoto model, a final estimate of $(\Omega_p, \gamma) = (0.918, 0.820)$, which differs by approximately $0.003$ in both components from the value obtained by the independent log-spiral method. This accuracy is substantially better than that achieved in previous implementations of the matrix method \citep{JH05}.

The log-spiral method achieves \texttt{x.xxxx} accuracy or better for the models considered here. However, the relationship between the expansion coefficients of the surface density and potential is fixed for zero softening, so the method cannot be extended to softened models. Its linear form (analogous to the standard linear matrix eigenvalue problem, 
$\omega\, \textsf{B\,x = A\,x}$) suffers from a numerical rank deficiency related to the Landau--Pollak--Slepian theorem \citep{Landau62, Slepian83}: the kernel matrix becomes ill-conditioned when the number of orbits exceeds the effective bandwidth $n_{\rm LPS}\approx \Delta u\,\Delta\alpha/\pi$, and the nonlinear form must be used instead. A detailed description of the log-spiral method will be given in a separate paper.

In contrast, the generalized linear method presented in this paper naturally incorporates gravitational softening \citep[Plummer softening is used in this paper, but the method can be generalized to other prescriptions, see, e.g.,][]{RFD19}, enabling the computation of eigenmodes for softened disk models. This is important for direct comparison with $N$-body simulations, which typically employ softened gravity. While less accurate than the log-spiral method, the level of accuracy attainable with the linear method is sufficient for studying the instability properties of stellar disks.


The implementation is openly available on GitHub,\footnote{\url{https://github.com/epolyach/pme_gpu_public}}
and a frozen version is archived on Zenodo \citep{pme_gpu_zenodo}.

\begin{acknowledgments}
This research was partially supported by the Russian Academy of Sciences Program No. 28 (subprogram II, 'Astrophysical Objects as Cosmic Laboratories) and the Ministry of Science and Higher Education of the Russian Federation (I. Shukhman). 
\end{acknowledgments}

\newpage

\bibliography{main}{}
\bibliographystyle{aasjournal}

\bigskip
\appendix

\section{Alternative Derivation of the Generalized Linear Matrix Equation}
\label{app:A}
Let the unperturbed DF be
 \be F_0(J,L)={\mathfrak F}(J,L)\,\Theta(L),
 \label{eq:F_0}
 \ee
  and the perturbed
 \be
 F(J,L,\bw)=[{\mathfrak F}(J,L)+f(J,L,\bw)]\Theta\Bigl(L+g(J,\bw)\Bigr),
 \ee
so the first order perturbation is
\be
    F-F_0\approx  F_1(J,L,\bw)\equiv f(J,L,\bw)\,\Theta(L)+{\mathfrak F}(J,0)\,g(J,\bw)\,\delta(L).
\ee

Here $\Theta(x)$ is the Heaviside step function and  $L+g(J,\bw)=0$ is equation of the perturbed boundary of the 4D phase space region $(J,L;w_1,w_2)$, which in the unperturbed state was $L=0$. We expand in harmonics $f(J,L,\bw)$ and $g(J,\bw)$: 
\be
 f(J,L,\bw)=e^{\i mw_2-\i\omega t}\sum\limits_{l=-\infty}^\infty f_l(J,L)\, e^{\i lw_1}
\ee
\be
g(J,\bw)=e^{\i mw_2-\i\omega t}\sum\limits_{l=-\infty}^\infty g_l(J)\, e^{\i lw_1}
\ee
so that
\be
{\cal F}_l(J,L)=f_l(J,L)\,\Theta(L)+{\mathfrak F}(J,0)\,g_l(J)\,\delta(L).
\label{eq:cal_F}
\ee
   Substitute (\ref{eq:cal_F}) into equation (\ref{eq:Zh2}) (see also \cite{PJ15})
\be
\Bigl[\omega-l\,\Omega_1(J,L)-m\,\Omega_2(J,L)\Bigr]\,{\cal F}_l=G \Bigl(l\,\frac{\p F_0}{\p J}+m\,\frac{\p F_0}{\p L}\Bigr)
\int dJ' dL'\sum\limits_{l'}\Pi_{ll'}(J,L;J',L')\,{\cal F}_{l'}(J',L').
\label{eq:Zh}
\ee
We write using (\ref{eq:F_0}) and (\ref{eq:cal_F}), 
\begin{multline}
\Bigl[\omega-l\,\Omega_1(J,L)-m\,\Omega_2(J,L)\Bigr]\,\Bigl[f_l(J,L)\Theta(L)+{\mathfrak F}(J,0)\,g_l(J)\,\delta(L)\Bigr]\\
=G \Big[\Bigl(l\,\frac{\p {\mathfrak F}}{\p J}+m\,\frac{\p {\mathfrak F}}{\p L}\Bigr)\,\Theta(L) +m\,{\mathfrak F}(J,0)\,\delta(L)\Bigr]\\
 \times \int dJ'\left[\int\limits_{L'>0} dL'\sum\limits_{l'}\Pi_{l l'}(J,L;J',L')\,f_{l'}(J',L')+ \sum\limits_{l'}\Pi_{l l'}(J,L;J',0)\,{\mathfrak F}(J',0)\,g_{l'}(J')\right].
\label{eq:full}
\end{multline}
We equate the expressions on the left and right sides of the equation, separately for $\Theta(L)$ and for $\delta(L)$. We obtain two equations for the functions $f_l(J,L)$ and $g_l(J)$
\[
\Bigl[\omega-l\Omega_1(J,L)-m\Omega_2(J,L)\Bigr]\,f_l(J,L)
\]
\be
=
G \,\Bigl(l\,\frac{\p {\mathfrak F}}{\p J}+m\,\frac{\p {\mathfrak F}}{\p L}\Bigr)\,
\int dJ'\left[\int\limits_{L'>0} dL'\sum\limits_{l'}\Pi_{l l'}(J,L;J',L')\,f_{l'}(J',L')+ \sum\limits_{l'}\Pi_{l l'}(J,L;J',0)\,{\mathfrak F}(J',0)\,g_{l'}(J')\right],
\ee

\[
\Bigl[\omega-l\Omega_1(J,0)-m\Omega_2(J,0)\Bigr]\,{\mathfrak F}(J,0)\,g_l(J)
\]
\be
= G\,m\,{\mathfrak F}(J,0)
  \int dJ'\left[\int\limits_{L'>0} dL'\sum\limits_{l'}\Pi_{l l'}(J,0;J',L')\,f_{l'}(J',L')+ \sum\limits_{l'}\Pi_{l l'}(J,0;J',0)\,{\mathfrak F}(J',0)\, g_{l'}(J')\right].
  \label{eq:2nd}
\ee
This system of two equations generalizes equation (\ref{eq:Zh}) to the case where ${\mathfrak F}(J,L=0)\ne 0$. In this case, in equation (\ref{eq:2nd}), we can divide both sides by ${\mathfrak F}(J,L=0)$ and obtain the final system:
\begin{multline}
\Bigl[\omega-l\Omega_1(J,L)-m\Omega_2(J,L)\Bigr]\,f_l(J,L)
 \\ =
G \,\Bigl(l\,\frac{\p {\mathfrak F}}{\p J}+m\,\frac{\p {\mathfrak F}}{\p L}\Bigr)\,
\int dJ'\left[\int\limits_{L'>0} dL'\sum\limits_{l'}\Pi_{l l'}(J,L;J',L')\,f_{l'}(J',L')+ \sum\limits_{l'}\Pi_{l l'}(J,L;J',0)\,{\mathfrak F}(J',0)\,g_{l'}(J')\right],
\label{eq:f_l}
\end{multline}

\begin{multline}
\Bigl[\omega-l\Omega_1(J,0)-m\Omega_2(J,0)\Bigr]\,g_l(J)\\
    =
G\,m\
\int dJ'\left[\int\limits_{L'>0} dL'\sum\limits_{l'}\Pi_{l l'}(J,0;J',L')\,f_{l'}(J',L')+ \sum\limits_{l'}\Pi_{l l'}(J,0;J',0)\,{\mathfrak F}(J',0)\, g_{l'}(J')\right].
\label{eq:g_l}
\end{multline}
If ${\mathfrak F}(J,L=0)=0$, the equation (\ref{eq:2nd}) turns into the identity $0=0$, and the equation (\ref{eq:f_l}) coincides with the usual equation (\ref{eq:Zh}). It is implied that in the l.h.s. of (\ref{eq:g_l}) $\Omega_2(J,0)=\frac{1}{2}\,\Omega_1(J,0)>0$.

The pair of equations (\ref{eq:f_l}) and (\ref{eq:g_l}) represent the final system of the generalized linear method.

\bigskip
It can be shown that this pair of equations is equivalent to the equation (\ref{eq:final}) on $h_l$, given in the main text. Let us put in (\ref{eq:f_l}) $L=0$. Comparing the resulting equation with ({\ref{eq:g_l}),
we find 
\be 
\Bigl[\omega-l\,\Omega_1(J,0)-m\,\Omega_2(J,0)\Bigr]\,f_l(J,0) 
= \Bigl[\omega-l\,\Omega_1(J,0)-m\,\Omega_2(J,0)\Bigr]\,\Bigl(l\,\frac{\p {\mathfrak F}}{\p J}+m\,\frac{\p {\mathfrak F}}{\p L}\Bigr)_{L=0}\frac{g_l(J)}{m}, 
\ee
or, dividing both sides by $\Bigl[\omega-l\,\Omega_1(J,0)-m\,\Omega_2(J,0)\Bigr]$,
\be
g_l(J)=\frac{m}{\Bigl(l\,\dfrac{\p {\mathfrak F}}{\p J}+m\,\dfrac{\p {\mathfrak F}}{\p L}\Bigr)_{L=0}}\,f_l(J,0).
\ee
It is convenient to introduce the variable
\be
h_l(J,L)=\frac{f_l(J,L)}{l\,\dfrac{\p {\mathfrak F}}{\p J}+m\,\dfrac{\p {\mathfrak F}}{\p L}}.
\ee
Then
\be
g_l(J)=m\,h_l(J,0).
\ee
Substituting $g_l(J)$ and $f_l(J,L)$ expressed through $h_l(J,0)$ and $h_l(J,L)$ into (\ref{eq:f_l}), 
we get the equation 
\begin{multline} 
\Bigl[\omega-l\Omega_1(J,L)-m\Omega_2(J,L)\Bigr]\,h_l(J,L)\\= 
\int dJ'\left[\int\limits_{L'>0} dL'\sum\limits_{l'}\Pi_{l l'}(J,L;J',L')\, \Bigl(l'\,\frac{\p {\mathfrak F}}{\p J'}+m\,\frac{\p {\mathfrak F}}{\p L'}\Bigr)\,h_{l'}(J',L')+\Pi_{ll'}(J,L;J',0)\, {\mathfrak F}(J',0)\,m\,h_{l'}(J',0)\right],
\label{eq:alternative} 
\end{multline} 
which coincides with (\ref{eq:final}).

\section{Grid variables and the Jacobian}
\label{app:B}
For models with softened central potentials, it is convenient to use $R_c = (r_{\rm min}+r_{\rm max})/2$, analogous to the semi-major axis of the Keplerian ellipse, and eccentricity $e=(r_{\rm max}-r_{\rm min})/(r_{\rm max}+r_{\rm min})$. The pericenter $r_{\rm min}(E,L)\equiv r_1(E,L)$ and apocenter $r_{\rm max}(E,L)\equiv r_2(E,L)$ of a star with energy $E$ and angular momentum $L$ are given by
\begin{equation}
r_1= R_c(1-e),\quad r_2=R_c(1+e),
\end{equation}
which are the roots of the equation
\begin{equation}\label{eq:turn_points}
2[E-\Phi_0(r)]r^2=L^2.
\end{equation}
We assume that the potential increases monotonically, $\Phi_0'(r)>0$, so that this equation has two real roots. However,  eccentricity parameterization has a drawback: the limiting radial orbits with $e\to 1$, $L>0$ and $e\to 1$, $L<0$ are separated. To address this issue, we introduce the {\it circularity} $\cc$ as
\begin{equation}
\cc = \sgn(L) (1-e).
\label{eq:circ}
\end{equation}
This parameterization has the advantage that both the $(E,L)$ and $(J,L)$ phase spaces map smoothly to $(R_c, \cc)$, which is important when studying disk models with a sharp cutoff near the radial orbits.

Let us calculate the Jacobian ${\cal J}$ of the transition from variables $(J,L)$ to $(R_c,\cc)$. To do this, we perform a chain of transitions
\be
(J,L)\to (E,L)\to (r_1,r_2)\to (R_c,e)\to (R_c,\cc).
\ee
The first and third transitions are trivial: $\d J\,\d L= {\d E\,\d L}/{\Omega_1}$, $\d r_1\,\d r_2 = 2R_c\,\d R_c\,\d e$.
Let us consider the second transition: $(E,L)\to (r_1,r_2)$. From (\ref{eq:turn_points}), we have
\be
L^2=2[E-\Phi_0(r_1)]\,r_1^2=
2[E-\Phi_0(r_2)]r_2^2,
\ee
and hence
\be
E=\frac{r_2^2\Phi_0(r_2)-r_1^2\Phi_0(r_1)}{r_2^2-r_1^2}<0,
\label{eq:E1}
\ee
We introduce
\be
    t_1=2[E(r_1,r_2)-\Phi_0(r_1)]\,r_1-r_1^2\Phi_0'(r_1),\quad t_2=2[E(r_1,r_2)-\Phi_0(r_2)]\,r_2-r_2^2\Phi_0'(r_2),
\ee
and then we have:
\be
\frac{\p E}{\p r_1}=\frac{t_1}{r_2^2-r_1^2},\ \ \frac{\p E}{\p r_2}=-\frac{t_2}{r_2^2-r_1^2},\quad
\frac{\p L^2}{\p r_1}=\frac{2t_1\,r_2^2}{r_2^2-r_1^2},\ \ \ \frac{\p L^2}{\p r_2}=-\frac{2t_2\,r_1^2}{r_2^2-r_1^2},
\ee
\be
\d J\,\d L=\frac{\d E \d L}{\Omega_1}=\frac{2R_c}{|L|\,\Omega_1}\,\frac{|t_1\,t_2|}{r_2^2-r_1^2}\,dR_c\,\d e.
\ee
From (\ref{eq:circ}), we have $e=1-|\cc|$ and finally
\be
{\cal J}=\Big|\frac{D(J,L)}{D(R_c,\cc)}\Big|=\frac{2R_c}{|L|\,\Omega_1}\,\frac{|t_1\,t_2|}{r_2^2-r_1^2}.
\ee
It can be shown that on radial orbits, $\cc=0$, the Jacobian is finite and equals
 \be
 {\cal J}_{\cc=0}=\frac{2R_c}{\Omega_1}\sqrt{2\,[\Phi_0(2R_c)-\Phi_0(0)]}\,\Phi_0'(2R_c),
 \ee
while on circular orbits it vanishes, ${\cal J}_{\cc=\pm 1}=0$.

\end{document}